\documentclass[acmsmall,nonacm]{acmart}

\usepackage{tabularx}
\usepackage{booktabs}
\usepackage{array}
\usepackage{placeins}
\usepackage{float}

\setcopyright{none}
\title{Same Team Label, Different Evidence: A Full-Text Audit of Claim Denominators in Human-AI Teaming Research}

\author{Hanjing Shi}
\email{hasa23@lehigh.edu}
\affiliation{%
  \institution{Lehigh University}
  \city{Bethlehem}
  \state{Pennsylvania}
  \country{USA}}

\author{Kimberly Wang}
\email{kim229@lehigh.edu}
\affiliation{%
  \institution{Lehigh University}
  \city{Bethlehem}
  \state{Pennsylvania}
  \country{USA}}

\author{Sabrina Doherty}
\email{sad328@lehigh.edu}
\affiliation{%
  \institution{Lehigh University}
  \city{Bethlehem}
  \state{Pennsylvania}
  \country{USA}}

\author{Dominic DiFranzo}
\email{djd219@lehigh.edu}
\affiliation{%
  \institution{Lehigh University}
  \city{Bethlehem}
  \state{Pennsylvania}
  \country{USA}}

\begin{abstract}
Human-AI Teaming (HAT) reviews often group studies by labels such as advisor, teammate, or coordinator. Yet the same label can describe one person taking AI advice, several people coordinating around AI, or a workflow that distributes authority and responsibility. Pooling these studies can therefore change the human unit behind a claim.

We examine how full-text evidence changes the set of studies behind a claim. We audited 86 full texts purposively selected from a 419-record title/abstract map. We find that full-text reading changed core membership for 40 records: 36 of 74 apparent core candidates moved out, while 4 of 12 boundary candidates moved in. Team vocabulary did not reliably identify the social unit: 14 of 27 human--AI dyads and 20 of 23 multi-human peer teams used team or collaboration terms. Only 20 of 86 papers specified who could see AI output. Four blinded language-model runs unanimously labeled 53 screening cases and 59 arrangements, yet 32\% and 34\% of those consensus decisions differed from the full-text labels.

These results identify claim-denominator drift as a synthesis problem in HAT research. We contribute a full-text audit centered on human arrangements and a claim-pooling checkpoint for deciding when evidence about trust, coordination, performance, efficiency, and accountability can be compared.
\end{abstract}



\begin{document}
\maketitle

\section{Introduction}

A literature review can be wrong even when every cited paper is real. The error can enter through the comparison itself. In Human-AI Teaming (HAT), the word ``team'' may refer to one person taking AI advice, a peer group using a chatbot, a classroom, a clinical workflow, a workplace delegation setting, or a governance process. Those settings all matter to HAI research. They do not produce the same kind of evidence.

Common HAT claims are relational. Trust may refer to one person's reliance on an AI, confidence among human teammates, or confidence in the institution that deploys the system. Performance may belong to one decision-maker, a group, or an organization. Collective intelligence requires information to be combined across members; accountability requires authority, traceability, and consequences. These outcomes cannot be inferred from the AI role alone.

One contrast makes the problem plain. In the audit below, an endoscopist reviewing lesion videos with AI support and a simulated ICU team coordinating with an AI agent are both relevant to HAT. The ICU study carries direct evidence about multi-human clinical coordination. The endoscopy study supports clinical AI-advice claims, but its contribution to clinical teamwork synthesis depends on whether the surrounding human arrangement is reported. If both are pooled under the same team label, the comparison quietly changes its human unit.

Existing reviews have progressively narrowed the broad HAT label. They clarify its definitions and conceptual vocabulary \citep{berretta2023defining,wang2026unpacking}, distinguish interaction patterns \citep{gomez2025collaboration}, map testbeds \citep{chung2026systematic}, and specify conditions for complementarity \citep{gonzalez2026science,hemmer2025complementarity}. Human-computation studies show how explanations and mental models affect complementary decisions \citep{bansal2019beyond,bansal2021whole}, while taxonomies of human and machine strengths describe possible divisions of labor \citep{rastogi2023taxonomy}. The newest HAT taxonomy classifies experimental studies by task interdependence, role, leadership, and communication structure \citep{hughes2026types}. These reviews classify what systems are, how people and AI exchange initiative, or which configurations and settings have been studied. The present audit addresses a prior synthesis decision: whether two papers that share a HAT label report comparable human relations, team processes, and outcome levels, and therefore belong in the same claim denominator.

This paper follows that connection through the published evidence. The analysis begins with the human arrangement behind an outcome: who acts, who has authority, who sees the AI output, how expertise is distributed, how stable the relation is, and what consequences are present. It then asks what the AI does inside that arrangement, how people are coupled to its output, which team process can occur, and where the reported outcome is located.

\textbf{RQ1: Arrangement recovery.} What human arrangements and evidence roles become visible through full-text audit of HAT and AI-supported collaboration studies?

\textbf{RQ2: Mechanism paths.} How do human and AI roles combine within an arrangement, through visibility, control, and responsibility, to organize different team processes?

\textbf{RQ3: Outcome alignment.} How do reported outcome and claim terms align with the human arrangement and relational coupling documented in each paper?

To answer these questions, the study combines a 419-record title/abstract map with an 86-paper full-text audit. The map establishes the vocabulary surrounding HAT; the audit determines what kind of evidence selected papers provide and which social relations their reports make recoverable. The paper makes three contributions. Empirically, it documents how full-text evidence changes claim denominators and how team vocabulary crosses distinct human arrangements. Conceptually, the Human-AI Teaming Mechanism Chain organizes human arrangement and AI role, relational coupling, team process, and outcome claim in one analytic sequence. Methodologically, a claim-pooling checkpoint turns that sequence into a procedure for deciding when trust, coordination, performance, and accountability evidence is comparable.

\section{Related Work: From AI Roles to Human Arrangements}

Review can reveal a mismatch between the unit a field studies and the phenomenon it seeks to explain. \citet{aghajari2023misinformation}, for example, compare a corpus of misinformation interventions with research on the drivers of misinformation. The comparison shows that intervention work centers individual content and individual responses even though prior beliefs and social context also shape how misinformation is encountered. Other reviews use classification to locate conceptual and methodological gaps: unstable definitions and measures of empathy \citep{genc2024empathy}; recurring forms of growing, appropriating, and coping with infrastructure, including its harmful effects \citep{lyu2025infrastructure}; and limited explicit definitions alongside varied frameworks for equity \citep{kim2025equity}. In each case, classification exposes a blind spot or neglected distinction and reorganizes the questions that follow.

Human-AI review papers use a similar pattern when they turn a broad label into a more precise analytic frame. \citet{shiyang2023designai} separate AI assisting designers, designers assisting AI, and broader designer-AI collaboration. \citet{zhang2025agencyreview} organize human-AI co-creation around agency patterns, operational control mechanisms, interaction contexts, and application domains. Their framework supports the classification and design of co-creative systems. Our unit of analysis is different: it begins with the reported human arrangement and asks what level of claim the paper can enter. Control is therefore one part of an evidence-admissibility check alongside authority, visibility, responsibility, team process, and outcome level, rather than the organizing lens of a design taxonomy. \citet{reza2025cowriting} combine a review of AI writing support with writer interviews to ask whether systems align with writing processes and user demands.

Recent HAT syntheses have moved even closer to the question of team form. \citet{schmutz2024aiteaming} foreground team constellations, trust, and shared cognition. \citet{kargarnovin2026testbeds} map 104 empirical HAT studies and describe a shift from controlled testbeds toward applied and higher-stakes settings. \citet{hughes2026types} apply psychological team taxonomies to 53 experimental HAT papers, derive five team types, and offer a checklist for specifying the team under study. A complementary line of theory models HAT as a sequence from inputs through mediators to outcomes; a recent critical-care IMOI model, for example, places team composition and interface design before trust, communication, coordination, and performance \citep{korentsides2026imoi}. The present audit asks whether published reports contain the relational evidence needed to instantiate such a sequence. Team taxonomy distinguishes configurations. The mechanism-chain analysis traces how a configuration makes a process possible and which outcome level the reported evidence can support.

That move from category to evidence draws on cooperative-work and organizational theory. Boundary objects explain how a shared artifact can coordinate communities without erasing local meanings \citep{star1989institutional}. Adaptive structuration theory places a technology's effects in the practices through which groups appropriate it \citep{desanctis1994capturing}. Common-ground theory explains how jointly maintained knowledge supports coordination \citep{clark1991grounding}, while transactive-memory and team-process research connects expertise and temporal structure to collective performance \citep{lewis2003measuring,marks2001temporally,cooke2013interactive}. Cooperative-work research further distinguishes awareness from the articulation work required to keep joint activity moving \citep{dourish1992awareness,schmidt1992taking}. Studies of invisible labor show that a clean system account can omit the work that sustains it \citep{star1999layers}. Sociotechnical analyses likewise show why accountability cannot be inferred from a system description alone: responsibility may be misassigned when control is distributed, and abstraction can exclude the social actors and contexts that shape an intervention \citep{elish2019moral,selbst2019fairness}.

Together, these traditions motivate the coding design that follows. Rather than accept ``team'' or an AI role as a stable explanatory unit, the audit records the relations through which a role can affect reliance, coordination, collective work, or accountability.

\section{Method: Corpus Construction and Full-Text Audit}

\subsection{Review design}

Search relevance and evidentiary use are answered at different levels of a paper. The title/abstract map identifies where HAT and AI-supported collaboration vocabulary appears. Full-text reading then determines what selected papers studied, which human relations surrounded the AI, and what outcomes those relations can support. The map therefore supplies the search frame and a record of surface similarity; evidence-role, arrangement, authority, visibility, stakes, reporting quality, and claim-scope findings come from the audit layer.

Methodological precedents for this sequence come from review designs that diagnose how a field organizes evidence. PRISMA and PRISMA-ScR provide the reporting-transparency baseline for systematic and scoping reviews \citep{page2021prisma,tricco2018prisma}. Scoping methods accommodate heterogeneous concepts and study designs \citep{arksey2005scoping,levac2010scoping}; systematic mapping separates a research landscape from later synthesis \citep{petersen2008systematic}. Thematic and critical interpretive synthesis then support a closer reading of how reported concepts become comparable evidence \citep{thomas2008thematic,dixonwoods2006critical}. In this study, the 86 papers form a purposive audit set designed to expose changes in comparison across varied arrangements.

\subsection{Corpus construction}

Corpus construction began with a 435-record title/abstract map because the literature is scattered across human-computer interaction, cooperative work, human factors, learning sciences, organizational studies, human-autonomy teaming, and human computation. The map combined targeted search strings, a seed bibliography, OpenAlex metadata expansion, and backward snowballing from review and taxonomy papers. The search map used two query families: Human-AI Teaming terms such as ``human-AI teaming,'' ``human-agent teaming,'' ``human-autonomy teaming,'' ``AI teammate,'' and ``collaborative AI''; and collaboration terms such as ``AI-supported collaboration,'' ``team,'' ``group,'' ``collaborative work,'' and ``coordination.'' At record level, the map retained matched query terms and source labels so that selection and later diagnostic checks could be traced back to title/abstract signals. Duplicate handling used DOI, title, version, and metadata checks, leaving 419 primary records.

Selection for the full-text audit focused on arrangement neighborhoods where comparison errors were most likely to matter: dyadic human-AI relations, multi-human peer teams, instructional groups, workplace and organizational workflows, expert-client or clinical workflows, and governance or oversight settings. Eligibility required three signals in the title, abstract, metadata, or seed-corpus status: a human-AI arrangement or plausible boundary case; a mechanism or outcome relevant to human computation and AI-supported collaboration, such as collaboration, delegation, trust, performance, accountability, learning, oversight, or decision quality; and enough access to support full-text evidence-location coding.

Fifty-seven retained records came from the seed audit corpus and had already passed the same title/abstract eligibility logic before OpenAlex expansion. Another 29 OpenAlex records filled underrepresented buckets. Before full-text adjudication, the 86-record set covered six target buckets: 24 multi-human peer-team records, 14 governance/oversight records, 14 workplace/organizational records, 12 dyadic human-AI records, 12 instructional records, and 10 expert-client/clinical records. It also included 74 core or near-core empirical candidates and 12 boundary or dyadic mechanism candidates. The remaining 333 primary records stay in the title/abstract map as search context: they preserve the broader vocabulary neighborhoods, while full-text claim labels are drawn from the 86 records that received evidence-location coding. Full-text coding then revised the initial labels; some selected records became boundary, background, or non-prevalence/context records after evidence-location review and lead adjudication.

\begin{table}[t]
\caption{Selection path from the title/abstract map to the full-text audit set.}
\label{tab:selection-flow}
\begin{tabularx}{\linewidth}{>{\raggedright\arraybackslash}p{0.20\linewidth}>{\raggedright\arraybackslash}p{0.10\linewidth}>{\raggedright\arraybackslash}p{0.32\linewidth}X}
\toprule
Stage & Records & Basis & Disposition \\
\midrule
Map construction & 435 & Targeted queries, seed bibliography, OpenAlex metadata expansion, and backward snowballing & Entered the title/abstract map. \\
Duplicate and version reconciliation & $-16$ & DOI, title, version, and metadata checks & 419 primary records remained. \\
Prior audit corpus retained & 57 & Met the title/abstract eligibility signals as core, near-core, or boundary candidates & Entered the full-text audit set. \\
Maximum-variation expansion & $+29$ & Filled underrepresented clinical (7), instructional (8), dyadic (1), governance (7), and workplace (6) targets & Entered the full-text audit set, bringing the six initial arrangement buckets to their planned coverage levels. \\
Map records not sampled for full-text audit & 333 & Selection stopped after the prior corpus and expansion records filled the planned variation targets & Remained in the map; no full-text evidence role, arrangement, or exclusion reason was assigned. \\
Full-text audit & 86 & Evidence-location coding, normalization, and adjudication & Supplied all claim-scope results reported in this paper. \\
\bottomrule
\end{tabularx}
\end{table}

This procedure leaves 333 records at the map layer. They preserve the search space around the audit but do not enter the later counts, because only the 86 sampled records received the evidence-location coding needed to determine claim scope.

At record level, the audit table preserves the path from source metadata to full-text judgment. Each row includes an internal record ID, bibliographic fields, source dataset, duplicate status, screening rationale, matched query terms, retrieval status, and full-text evidence notes. Record IDs such as OA043, P19, and P22 are audit identifiers that link manuscript examples to coding rows; they are not external citation keys.
\FloatBarrier

\subsection{Evidence roles and codebook development}

Each row then receives an evidence-role label. Core empirical records involve a recoverable human-AI arrangement and report evidence relevant to interaction, collaboration, responsibility, trust, performance, participation, coordination, learning, oversight, or decision quality. Observed, experimental, and deployed studies provide direct interaction evidence. Design-study records enter the core set only when they report participant or workflow evidence tied to a concrete arrangement. Boundary records are used for mechanism or role framing when the paper informs the concept but does not report a concrete arrangement that can carry the focal claim. Common cases include one-human plus AI studies that inform reliance or attribution, future-facing scenario studies, design frameworks without observed team interaction, and role/taxonomy papers that help define AI configuration. Background records include reviews, conceptual frameworks, and theory sources used for framing.

Codebook development began with the synthesis problem and the literatures reviewed above. Team and organization research informed fields for expertise coordination, temporal structure, team processes, and task context \citep{lewis2003measuring,marks2001temporally,salas2005big,cooke2013interactive}. Calibration added explicit fields for roles and authority because these distinctions changed the comparison unit. Human-automation research informed fields for control, automation level, and reliance \citep{lee2004trust,parasuraman2000model}. Communication and cooperative-work research informed fields for visibility, common ground, coordination, and articulation work \citep{clark1991grounding,dourish1992awareness,schmidt1992taking}. Accountability and sociotechnical AI research informed fields for responsibility and social context \citep{elish2019moral,selbst2019fairness}. The codebook also required an evidence location for every interpretive field. Calibration then turned these dimensions into coding gates because records that looked similar at the title/abstract level often split apart during full-text reading: some reported dyads, some reported scenarios, and some reported application domains without actual study stakes.

In the main text, five gates summarize the codebook. The screening gate assigns each paper to core, boundary, background, exclusion, or non-prevalence/context status. The evidence-role gate records whether the record supports core empirical claims, mechanism interpretation, background theory, or exclusion. The arrangement gate codes the focal human relation into dyad, multi-human peer team, instructional group, workplace/organizational workflow, expert-client/clinical workflow, governance/oversight setting, background/theory, unclear, or low relevance. The authority and stakes gates distinguish peer, hierarchical, instructional, expert-client, governance, mixed, and unclear authority from low-stakes lab, graded/instructional, workplace, clinical/professional, safety-critical, governance/platform, and unclear stakes. The reporting-quality gate records whether the arrangement is explicit, partial, implicit, or not recoverable, and each included or boundary record provides a quote, section, page, table, or other location cue.

\begin{table}[t]
\caption{Operational codebook gates used in the reporting audit.}
\label{tab:codebook}
\begin{tabularx}{\linewidth}{>{\raggedright\arraybackslash}p{0.20\linewidth}>{\raggedright\arraybackslash}p{0.28\linewidth}>{\raggedright\arraybackslash}p{0.24\linewidth}X}
\toprule
Gate & Definition & Controlled codes & Decision rule \\
\midrule
Evidence role & Kind of claim the record can support in this audit & Core empirical; boundary mechanism; background theory; non-prevalence/context & Count a record as core when the full text reports direct interaction, deployment, experiment, or design-study evidence tied to a recoverable human-AI arrangement. \\
Team form & Focal human relation around the AI & Dyad; multi-human peer team; instructional group; workplace workflow; expert-client/clinical workflow; governance/oversight; unclear; low relevance & Code the observed focal relation over the paper title, preferred label, or application domain. \\
Authority & Relation of control, delegation, or override & Peer; hierarchical; instructional; expert-client; governance; mixed; unclear & Use the reported control relation: who can act on, contest, or override AI output. \\
Stakes & Consequences present in the study design & Low-stakes lab; graded/instructional; workplace; clinical/professional; safety-critical; governance/platform; unclear & Code consequences present in the study design; treat speculative downstream uses as discussion context. \\
Reporting quality & Recoverability of the human arrangement & Explicit; partial; implicit; not recoverable & Require a quote, page, section, table, or equivalent location cue for included and boundary records. \\
\bottomrule
\end{tabularx}
\end{table}

\subsection{Coding workflow}

For each retained record, trained coders recorded bibliographic information, retrieval status, screening decision, evidence role, and the human arrangement described in the text. The arrangement fields cover human count and roles, authority, expertise, temporal stability, communication mode, task domain, accountability stakes, AI role, AI visibility, AI autonomy/control, responsibility model, outcomes, attribution or accountability measurement, reporting quality, and theorization. Each included or boundary record requires a quote, page, section, table, or other location cue. Mechanism memoing identifies the human relation assumed by the study, the AI visibility or control arrangement, the social mechanism invoked or implied, and the scoped outcome claim.

Synthesis was decomposed into auditable tasks: record retrieval, controlled-field coding, evidence-location capture, mechanism memoing, schema normalization, agreement diagnostics, and adjudication. It is informed by human-computation work on dividing and checking literature-screening judgment \citep{krivosheev2017crowdsourcing}. Coders first received the workflow memo, codebook, decision rules, and calibration sheet. Twelve calibration records were coded independently by all three coders before discussion. Seventy-four additional records received a primary coder, and 12 of those received secondary overlap coding. Thus, 24 of 86 records had overlapping coder entries, and the remaining records contributed primary coding plus evidence-location notes. In total, the assignment packet created 122 record-coding tasks: Coder 1 had 41 tasks, Coder 2 had 41, and Coder 3 had 40.

\subsection{Mechanism-chain synthesis}

The final analytic pass connected fields that are often reported separately. For each core or boundary record, the analysis read five elements in sequence: the human arrangement and the roles people occupied within it; the AI role; the relational coupling created by visibility, autonomy, control, and responsibility; the team process named or implied in the mechanism memo; and the level of the reported outcome. Evidence locations and the finalized arrangement, authority, and stakes codes constrained each interpretation.

Outcome level was coded from the \texttt{outcomes\_reported} field, which contains outcome and claim terms extracted during full-text coding. A keyword rule assigned the recorded terms to one or more of four levels. Individual covers reliance, calibration, confidence, mental models, and a single decision-maker's task result. Team covers team performance, transactive memory, speaking up, coordination, cohesion, shared cognition, and participation. Workflow covers efficiency, throughput, adoption, delegation, handoff, and organizational division of labor. Institutional covers accountability, responsibility attribution, governance, oversight, and contestation. Coding is multi-label because papers commonly use terms at several levels. Records whose outcome field read ``not recoverable from available full text'' were excluded from this analysis, leaving 78 of 86. Because the rule operates on coder-extracted reporting vocabulary, it does not establish that every matched construct was independently operationalized or measured. The complete keyword list and record-level assignments accompany the analysis scripts.

Comparing these sequences produced five recurring analytic routes: advisory calibration, shared coordination, task orchestration, collective facilitation, and supervisory accountability. The routes were synthesized qualitatively from coded sequences and mechanism memos rather than assigned as mutually exclusive paper types. A system may advise one person while also exposing a trace to a supervisor, and the two relations support different claims. The routes connect recurring combinations in the audit to prior theory; they do not estimate causal effects or the prevalence of each mechanism in HAT research.

\subsection{Normalization, agreement diagnostics, and adjudication}

After human coding, a scripted normalization audit ran before agreement was computed. The audit trimmed whitespace, resolved spelling and slash variants, and mapped obvious codebook aliases to the controlled vocabulary. The script did not assign final codes or make substantive recoding decisions. It preserved raw coder entries, wrote companion normalized values and flags, and routed any interpretation-changing value to lead review. Substantive cases stayed in the adjudication queue.

Agreement marked fields where coders needed interpretive judgment. This choice follows qualitative-methods guidance in human-computer interaction and computer-supported cooperative work, where inter-rater reliability is useful when it matches the analytic goal and coding design \citep{mcdonald2019reliability}. Most records in this audit received primary coding, with planned overlap for calibration and breakpoint detection. The overlap statistics function as workflow diagnostics: they identify fields where interpretation entered the coding process. Pre-adjudication agreement is computed on independent, comparable coder entries; finalized screening codes are used for denominator analysis.

Lead adjudication produced the final analytic coding used in this manuscript through rule-based synthesis. The audit retained the raw coder values, normalized candidates, issue type, final code, reason, and any codebook-rule change for each queued issue. Lead adjudication assigned final codes and reasons to all 171 queued field-level issues and all 141 record-level final cells flagged by the worklist. Some issues were mechanical schema errors. Others reflected codebook ambiguity or reporting ambiguity and were kept visible as coding breakpoints.

\subsection{Surface-triage diagnostics}

After coding, two diagnostics examined whether the published surface reproduces the distinctions found through full-text reading. Both were scored against the adjudicated full-text labels, which were held in a separate file throughout.

The first is a TF--IDF diagnostic. For each audited record, the script concatenated title, abstract, screening rationale, and matched query terms; built unigram and bigram vectors; clustered records with cosine k-means for $k=5$ to $12$; selected $k$ by silhouette; and ran leave-one-out nearest-centroid triage against the final screening decision.

Because a weak representation and an underdetermined surface can produce the same result, the diagnostic was repeated with Claude Opus 5 and GPT-5.6 Sol, accessed through their chat interfaces on August 20, 2026. Each model was run under two blinded conditions using the default interface settings; the interfaces exposed no sampling controls. Condition A supplied the same fields the TF--IDF diagnostic received, making the two directly comparable. Condition B supplied only title, venue, year, and abstract, which is what the published surface shows a reader. Every run received the controlled label definitions and was asked to assign a screening decision, an arrangement bucket, and a self-reported confidence level for all 86 records. Abstracts were retrievable for 40 of 86 records; the remainder supplied title and venue only. No run had access to the full texts, the coding sheets, the adjudication trail, or any other run's output, and each condition was executed in a separate session. Each condition was run once per model. These runs are error analyses of the surface representation, not evaluations of general model capability; the adjudicated full-text codes remained the analytic labels.

\subsection{Analytic materials}

Analytic materials consist of published scholarly records, bibliographic metadata, and the research team's coding work product, reported through functional labels such as Coder 1--Coder 3. No new human-subject interaction data were collected. Copyrighted full texts are not redistributed; manuscript examples use bibliographic identifiers, aggregate codes, and evidence-location cues. Appendix~\ref{app:evidence-map} provides a record-level map of which analytic fields were recoverable, while the machine-readable audit table retains record IDs, full code values, and source-location notes.

Results move from evidence role to arrangement, mechanism path, and outcome level. Table~\ref{tab:codebook} defines the coding gates; the findings report how records moved across analytic roles and how the retained evidence connects AI roles to team processes.

\section{Findings}

\subsection{Full-text audit changed the claim denominators}

Full-text coding divided records that looked similar during screening. Of the 74 records initially treated as core or near-core empirical candidates, 38 remained core, 16 moved to boundary-mechanism evidence, 12 moved to background or theory, and eight remained context records because the focal arrangement was not recoverable. Movement also went in the other direction: four of the 12 initial boundary or dyadic-mechanism candidates became core, seven remained boundary records, and one became background. The final evidence base therefore comprised 42 core empirical records, 23 boundary-mechanism records, 13 background/theory records, and eight context records. In all, 36 of the 74 apparent core candidates left the core denominator after full-text reading.

The final arrangements cut across the categories used to construct the sample. The audit identified 27 human--AI dyads, 23 multi-human peer teams, eight workplace or organizational workflows, five expert-client or clinical workflows, two governance settings, and two instructional groups. Ten records were background reviews or taxonomies, eight remained unclear after full-text screening, and one had no recoverable focal arrangement. Reporting quality tracked evidentiary use: 63 records stated the arrangement explicitly, including all 42 core empirical records, whereas all eight context records were not recoverable.

\begin{figure}[H]
\centering
\includegraphics[width=\linewidth]{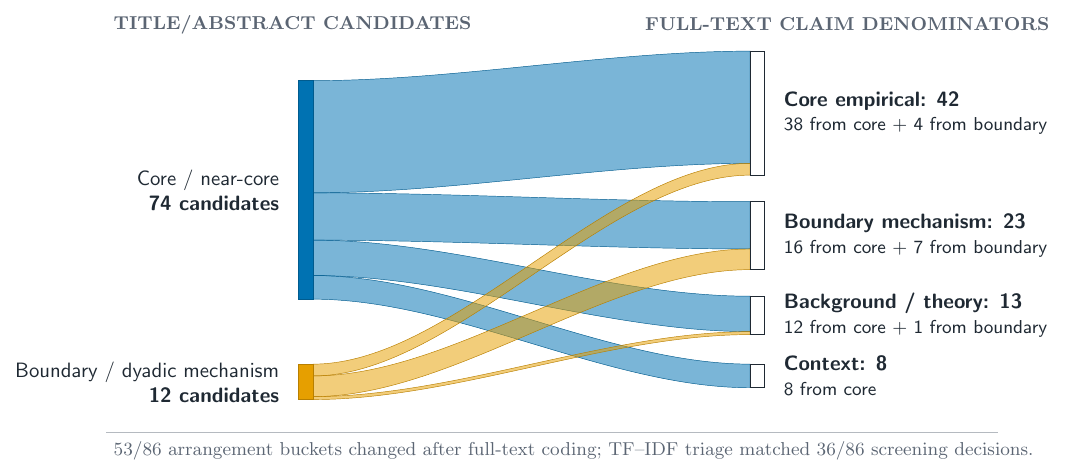}
\caption{Movement from title/abstract candidates to full-text claim denominators. Blue links originate from the 74 core/near-core candidates; orange links originate from the 12 boundary/dyadic-mechanism candidates. Link width is proportional to record count. Full-text reading moved 36 apparent core candidates out of the core empirical set and moved four boundary candidates into it.}
\Description{A flow diagram with two title-and-abstract source groups and four full-text evidence roles. Of 74 core or near-core candidates, 38 became core empirical, 16 boundary mechanism, 12 background theory, and eight context records. Of 12 boundary or dyadic candidates, four became core, seven boundary, and one background. The final totals are 42 core, 23 boundary, 13 background, and eight context records. A note reports that 53 of 86 arrangement buckets changed and TF-IDF triage matched 36 of 86 screening decisions.}
\label{fig:denominator-status}
\end{figure}

Figure~\ref{fig:denominator-status} shows how title/abstract relevance became a claim-specific evidence base. Dyadic studies can directly support conclusions about advice, reliance, calibration, or human--AI task performance. They become evidence about multi-human coordination only when the surrounding human relation is observed and reported.

\subsection{Team vocabulary crossed different human arrangements}

One recurring breakpoint was the use of team language for one human plus one AI. Such papers can provide strong evidence about reliance, attribution, calibration, or AI-mediated communication. Evidence about multi-human coordination requires another relation: people coordinating with one another around the AI. The adjudication rule therefore followed the arrangement enacted in the study rather than the label used in its title.

Three records illustrate the consequence. OA043, ``Experimental evidence of effective human-AI collaboration in medical decision-making,'' reports endoscopists reviewing lesion videos with and without AI support \citep{reverberi2022experimental}. It supports clinical AI-advice claims, but not direct claims about coordination among clinicians. OA076 reports physicians and nurses working in simulated four-person ICU teams with an AI agent, so transactive memory and speaking up occur in an observed human--human arrangement \citep{bienefeld2023human}. P22, ``Is the Most Accurate AI the Best Teammate?,'' models a single human overseer and is used here to explain an optimization mechanism rather than multi-human teamwork \citep{bansal2021most}. The same vocabulary thus enters evidence about advice-taking, team coordination, and system design.

The outcome fields reproduce this overlap. Among 78 records with recoverable outcome or claim terms, 14 of the 27 dyadic records used at least one team-level term such as team performance, teamwork, collaboration, or coordination. Twenty of the 23 multi-human peer-team records did so. Table~\ref{tab:arrangement-outcome} shows the focal comparison. A review that selects on team outcome vocabulary alone would pool these arrangements even though only one contains a human--human relation. Denominator drift is therefore present in the vocabulary inherited from primary studies, not only in a reviewer's later inclusion decisions.

\begin{table}[H]
\caption{Reported outcome and claim vocabulary in the two focal arrangements. Coding is multi-label, and percentages are within arrangement.}
\label{tab:arrangement-outcome}
\small
\begin{tabularx}{\linewidth}{Xrrrrr}
\toprule
Arrangement & $n$ & Individual & Team & Workflow & Institutional \\
\midrule
Human--AI dyad & 27 & 19 (70\%) & 14 (52\%) & 4 (15\%) & 3 (11\%) \\
Multi-human peer team & 23 & 7 (30\%) & 20 (87\%) & 8 (35\%) & 4 (17\%) \\
\bottomrule
\end{tabularx}
\end{table}

\subsection{Relational coupling specified different mechanisms}

The AI role became more interpretable when read together with control and responsibility. Records coded as advisory most often carried individual-level vocabulary (19 of 25), while facilitation records most often carried team-level vocabulary (seven of eight). Responsibility showed a parallel descriptive pattern: where one human retained the final decision, 18 of 27 records used individual-level terms; where responsibility was shared by a team, 23 of 29 used team-level terms. These are overlaps in reported vocabulary rather than causal effects, but they show that relational coupling distinguishes evidence more directly than the teammate label alone.

Visibility could not be analyzed in the same way because most reports did not identify an audience. Only 20 of 86 records used an audience-specific value---private, shared, translucent cue, or hidden. The other 66 recorded a visible agent (45), unclear visibility (13), mixed visibility (six), or an embedded system (two). Those values establish that AI was present or perceivable, not whether its output was available to one participant, a group, or an evaluator. This is a reporting boundary, not evidence that visibility has no effect on team process.

Reading arrangement, AI role, control, visibility, responsibility, process, and outcome together produced five recurring analytic routes. Table~\ref{tab:claim-scope-examples} summarizes them without treating them as exclusive paper types or prevalence categories. Private advice can organize reliance and calibration. Shared output can become common ground when members can respond to it and recognize one another's responses \citep{clark1991grounding}. Allocation authority can organize delegation and handoffs. Generative material can enter collective work when members select, revise, and combine it. A visible trace can support accountability when it crosses an authority boundary and an actor can question, correct, or justify the resulting action.

\begin{table}[H]
\caption{Mechanism paths synthesized from coded relations and audit breakpoints. Paths are reusable analytic routes rather than exclusive team types or causal estimates.}
\label{tab:claim-scope-examples}
\small
\begin{tabularx}{\linewidth}{>{\raggedright\arraybackslash}p{0.16\linewidth}>{\raggedright\arraybackslash}p{0.22\linewidth}>{\raggedright\arraybackslash}p{0.27\linewidth}X}
\toprule
Mechanism path & Human and AI roles & Relational coupling and process & Outcome claim supported \\
\midrule
Advisory calibration & Human decision-maker; AI advisor & Private advice and retained human control organize reliance, override, and calibration & Individual trust in AI and decision quality \\
Shared coordination & Peer or professional team; AI teammate or shared resource & Shared output enters common ground, transactive memory, speaking up, and mutual adjustment & Team coordination, team trust, and collective decision quality \\
Task orchestration & Workers, experts, or managers; AI coordinator & Allocation and control rules organize delegation, handoff, workload, and self-efficacy & Workflow efficiency, satisfaction, and task performance \\
Collective facilitation & Contributing group members; AI facilitator & Members select, revise, and combine AI contributions through participation and knowledge sharing & Collective creativity, learning, or group performance \\
Supervisory accountability & Operator, subject, or evaluator; AI monitor or trace generator & Visibility across an authority boundary organizes monitoring, contestation, attribution, and responsibility & Procedural or institutional trust and accountability \\
\bottomrule
\end{tabularx}
\end{table}

Applied to the clinical contrast, these routes change the conclusion rather than merely adding context. The audited records do not form one body of evidence that AI improves clinical teamwork. They form distinguishable bodies of evidence about clinical advice, simulated team coordination, and oversight. Comparisons can proceed within a route, or across routes when the change in human arrangement is itself part of the explanation.

\subsection{Surface triage reproduced the same collapse}

The surface diagnostics asked whether titles, abstracts, and metadata recover the full-text distinctions. TF--IDF leave-one-out triage matched 36 of 86 screening decisions, below the 42 of 86 majority-class baseline. The language-model results in Table~\ref{tab:triage} were stronger but model-dependent. Claude Opus 5 matched 55 and 58 screening decisions under conditions A and B; GPT-5.6 Sol matched 45 and 47. Arrangement accuracy was more stable, ranging from 48 to 50 of 86 across all four runs. Supplying the screening rationale and matched query terms in condition A did not improve either model over the title/venue/year/abstract condition B.

\begin{table}[H]
\caption{Blinded surface-triage diagnostics against adjudicated full-text labels. Condition A supplies title, abstract, screening rationale, and matched terms; condition B supplies title, venue, year, and abstract.}
\label{tab:triage}
\small
\begin{tabularx}{\linewidth}{Xrr}
\toprule
Run & Screening decision & Arrangement bucket \\
\midrule
Majority class & 42/86 & 27/86 \\
TF--IDF leave-one-out nearest centroid & 36/86 & --- \\
Claude Opus 5, condition A & 55/86 & 49/86 \\
Claude Opus 5, condition B & 58/86 & 50/86 \\
GPT-5.6 Sol, condition A & 45/86 & 48/86 \\
GPT-5.6 Sol, condition B & 47/86 & 49/86 \\
\bottomrule
\end{tabularx}
\end{table}

Accuracy varied, but the disagreement structure repeated. Every one of the six run pairs agreed with one another on more records than either run matched the full-text labels. Within a model family, the two conditions agreed on 78 of 86 screening decisions and 76 arrangements for Claude, and 75 screening decisions and 73 arrangements for GPT. Across families, pairwise agreement remained 60--62 on screening and 64--69 on arrangement. Repeating the judgment or switching model families therefore did not make agreement a correctness signal.

Four-run consensus makes this point directly. All runs gave the same screening label for 53 records, but 17 of those decisions (32\%) diverged from full-text adjudication. They gave the same arrangement label for 59 records, but 20 (34\%) diverged. The shared pattern also concerned the audit's focal distinction. Claude recovered 13 of 23 multi-human peer teams in each condition, compared with 19 and 20 of 27 dyads. GPT recovered 11 of 23 multi-human teams in each condition, compared with 21 of 27 dyads. Five core records were read as boundary evidence by all four runs, and none of the eight context records was identified as pending. GPT also assigned high confidence to 44 and 40 arrangement decisions while matching 55\% and 52\% of them, below its own medium-confidence accuracy in both conditions.

The diagnostics do not establish why a particular model diverged. They show that, in this audit, surface-level consensus reproduced a stable simplification of the evidence base and under-recovered the human--human relation. Surface triage can prioritize reading; it cannot replace the full-text step that fixes the claim denominator.

\subsection{Reporting and adjudication preserved claim scope}

Reporting quality determined how a record could be used. An explicit report makes the arrangement visible in the paper body. The clinical decision-support study by \citet{jacobs2021designing}, for example, reports interviews and focus groups around time-constrained antidepressant decisions, allowing recovery of expert-client authority, clinical stakes, and the intended workflow. A partial report provides some social structure but leaves the enacted arrangement underspecified. \citet{song2024humanai} theorize collaboration roles and design conditions and therefore support mechanism interpretation, not observed team-process claims. A not-recoverable report stops the coding route rather than inviting the application domain to stand in for an observed relation.

\begin{figure}[H]
\centering
\includegraphics[width=\linewidth]{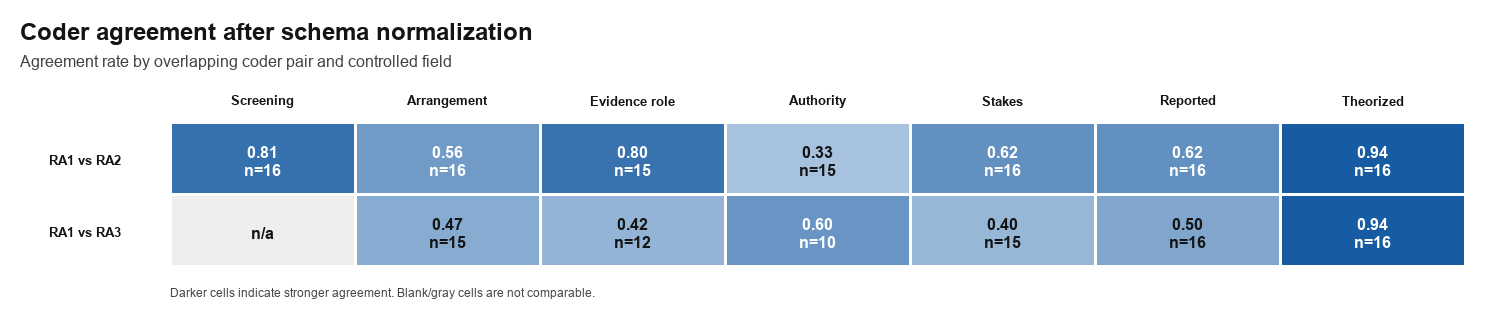}
\caption{Coder-overlap diagnostics before lead adjudication, after schema normalization. Cells show normalized agreement and comparable overlap size for each coder pair and controlled field. The gray cell marks a screening field without two comparable independent screening decisions; those records received final screening status through lead adjudication.}
\Description{A heatmap showing agreement rates for Coder 1 versus Coder 2 and Coder 1 versus Coder 3 across screening, arrangement, evidence role, authority, stakes, reporting, and theorization fields. Agreement is strongest for screening, evidence role, and theorization, and weaker for arrangement, authority, stakes, and reporting fields.}
\label{fig:agreement-heatmap}
\end{figure}

Adjudication separated mechanical variants, codebook mismatches, and interpretive ambiguity. Across 171 field-level issues, 73 were normalized coder disagreements, 40 were incomplete screening fields routed to adjudication, 38 were invalid or substantive values requiring review, and 20 were semantic aliases. Figure~\ref{fig:agreement-heatmap} reports comparable overlap after normalization. Coder 1--Coder 2 agreed on 13/16 screening decisions, 12/15 evidence-role labels, and 9/16 arrangement buckets. Comparable Coder 1--Coder 3 entries agreed on 7/15 arrangement buckets and 5/12 evidence-role labels. These diagnostics located decisions that required a textual evidence location and final rule.

\raggedbottom

Final rules kept topic, mechanism, and observed relation separate. Design studies describing future workflows remained useful for mechanism framing without becoming evidence of observed teamwork. One-human AI studies entered dyadic analyses, whereas multi-human claims required people to coordinate with one another around the AI. Clinical, workplace, and governance stakes were coded only when responsibility, grading, safety, oversight, or institutional consequences appeared in the study design. This separation lets a paper remain relevant without forcing all relevant papers into the same empirical denominator.

\section{The HAT Mechanism Chain and Claim-Pooling Checkpoint}

\begin{figure}[t]
\centering
\includegraphics[width=\linewidth]{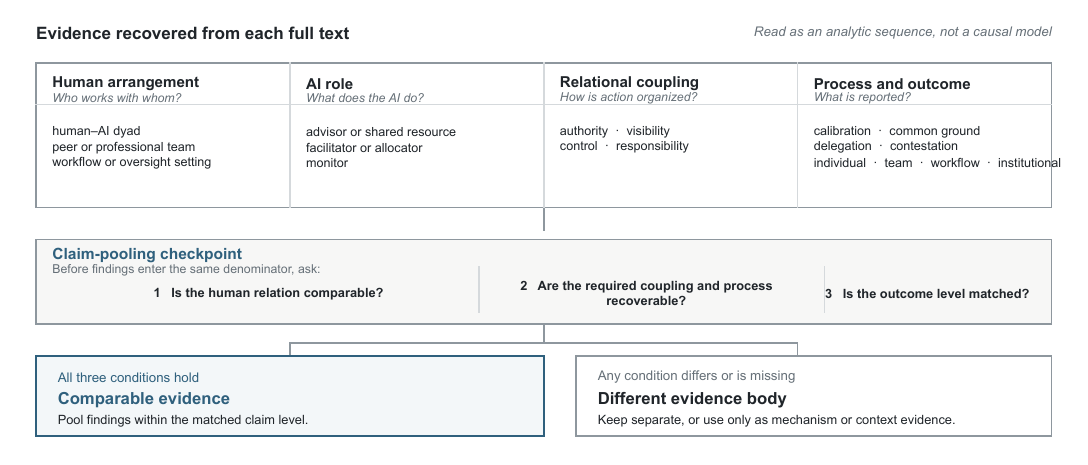}
\caption{The Human-AI Teaming Mechanism Chain as a claim-pooling procedure. For each full text, the reviewer records the human arrangement, AI role, relational coupling, team process, and outcome level. Findings enter the same denominator only when the human relation is comparable, the coupling and process required by the claim are recoverable, and the outcome level is matched. Otherwise, records remain separate evidence bodies or contribute only mechanism or context evidence. The sequence scopes evidence; it is not a causal model.}
\Description{A three-stage synthesis procedure. The first stage records four aligned fields from each full text: human arrangement, AI role, relational coupling, and process and outcome. The second stage asks whether the human relation is comparable, whether the required coupling and process are recoverable, and whether the outcome level is matched. If all three conditions hold, findings can be pooled within the matched claim level; if any condition differs or is missing, the records remain separate or are used only as mechanism or context evidence.}
\label{fig:mechanism-chain}
\end{figure}

Figure~\ref{fig:mechanism-chain} turns the audit dimensions into a decision procedure for evidence synthesis. A reviewer first recovers the human arrangement, AI role, relational coupling, team process, and outcome level from each full text. The claim-pooling checkpoint then asks whether the human relation is comparable, the coupling and process required by the claim are recoverable, and the outcome level is matched. The sequence determines whether findings can enter the same denominator; it does not estimate a causal model.

Relevance screening establishes that a paper concerns HAT; it does not establish that its findings belong in a particular comparison. The claim-pooling checkpoint operationalizes the chain. Before pooling a trust, performance, accountability, or complementarity result, the reviewer records the social conditions under which it was produced.

\begin{table}[H]
\caption{HAT Claim-Pooling Checkpoint adapted from prior theory and illustrated by audit breakpoints.}
\label{tab:synthesis-gate}
\small
\begin{tabularx}{\linewidth}{>{\raggedright\arraybackslash}p{0.18\linewidth}>{\raggedright\arraybackslash}p{0.32\linewidth}>{\raggedright\arraybackslash}p{0.23\linewidth}X}
\toprule
Dimension & Reporting question & Minimum detail & Claim-scoping consequence \\
\midrule
Authority & Who can accept, reject, override, or delegate AI output? & Peer, hierarchical, expert-client, instructional, governance, or unclear & Bounds delegation and control claims. \\
Expertise & Who is expected to know what? & Novice, expert, mixed, distributed, or unspecified expertise & Bounds trust, calibration, and reliance claims. \\
Temporal stability & Is the relation one-shot, short-term, repeated, or organizational? & Duration or recurrence of the human relation & Bounds claims about learning, routines, and transactive memory. \\
Visibility & Who sees AI output and human responses? & Private, shared, role-specific, public, or unclear visibility & Bounds coordination and common-ground claims. \\
Stakes & What consequences exist in the study design? & Lab, graded, workplace, clinical, governance, safety-critical, or unclear stakes & Bounds accountability and responsibility claims. \\
\bottomrule
\end{tabularx}
\end{table}

Taken together, the five questions locate a finding on the mechanism chain. Two studies can be pooled when the relation and process required by the outcome remain comparable. When only the process is shared, one record may explain a mechanism without serving as direct evidence for the same team-level outcome.

Authority gives the checkpoint its theoretical force. The same visibility design may support awareness in a peer team but become monitoring infrastructure in an instructional or managerial hierarchy. \citet{shi2026gets} treat AI disclosure as an epistemic-coordination problem in human-AI peer teams. In a different relation, \citet{feng2025group} examine student interaction with a generative-AI chatbot during collaborative problem solving. Comparing these settings requires asking what the visible trace permits participants to know and do: it may become peer-visible provenance, evidence for collaborative learning, or a record available to someone with evaluative authority.

\FloatBarrier
\flushbottom

\section{Discussion: From Team Configuration to Team Mechanism}

HAT research now has increasingly precise accounts of definitions, interaction patterns, testbeds, team types, and complementarity conditions \citep{gomez2025collaboration,chung2026systematic,wang2026unpacking,hughes2026types,gonzalez2026science}. The contribution here is not another classification of AI agency or roles. It specifies when a reported configuration provides evidence for a team-process claim and when it does not. A configuration identifies who or what belongs to a team; the mechanism chain asks what that arrangement allows participants to know, do, contest, and be held responsible for, then locates the outcome at the individual, team, workflow, or institutional level. Private advice, a shared object for coordination, and a trace available to an evaluator may involve the same AI system while producing different evidence.

\subsection{AI roles acquire meaning through human relations}

An AI advisor is not a stable social role. A private recommendation to one decision-maker enters a calibration process. The same recommendation shared among peers can become material for common ground, while a record visible to a manager can become monitoring evidence. Analysis centered on the human arrangement captures this change before an outcome is named. It also explains why adding social detail after the analysis is too late: authority and visibility participate in the mechanism that produced the result.

The audit also indicates which of these conditions the literature currently makes available. Control and responsibility were reported specifically enough to show descriptive alignment with outcome vocabulary: advisory and retained-decision arrangements more often carried individual terms, while facilitation and shared responsibility more often carried team terms. Visibility was not. Most records state that an AI was perceivable without stating to whom, so the condition that theory treats as central to common ground is the one the published record specifies least. The order of the chain is therefore an analytic commitment rather than a ranking by evidential strength, and closing the gap requires reporting practice to change rather than more re-reading of existing papers.

This account extends input--mediator--outcome models by making the reporting path itself inspectable. The critical-care IMOI model places trust, communication, and coordination between team inputs and performance \citep{korentsides2026imoi}. The mechanism chain identifies the evidence needed to instantiate that middle layer in a published study. It can therefore travel across domains without assuming that the mediator remains unchanged: shared output may support transactive memory in a clinical team, delegation may shape self-efficacy in task allocation, and visible moderation may alter institutional trust.

\subsection{Trust, collective performance, and accountability occupy different levels}

Trust is especially vulnerable to level confusion. Research on trust in AI commonly measures a person's cognitive or affective orientation toward a system \citep{glikson2020humantrust}. Team trust also develops through relations among members and can spread through communication and observed behavior \citep{zhou2026spread}. A HAT study should therefore identify the referent: trust in the AI, trust between human teammates, trust in the mixed team, or trust in the organization using the system. These constructs may interact, but one does not stand in for the others.

Collective performance requires a similar distinction. A person and an AI can outperform either component without demonstrating that several humans integrated knowledge as a team. Claims about collective intelligence need evidence of contribution integration, shared cognition, or coordinated action. Efficiency claims need the unit and cost to be visible: faster individual decisions, smoother handoffs, lower coordination burden, and higher organizational throughput are different outcomes. The mechanism chain keeps those measures attached to the process that produced them.

Accountability enters when action is connected to authority and consequences. Logging an AI output does not create accountability by itself. The trace must be available to an actor who can question, sanction, correct, or justify a decision, and the study must specify who retains control. This is why the same visibility feature may support awareness among peers yet increase surveillance or shift blame in a hierarchy. Treating accountability as a mechanism path, rather than an application-domain label, makes these differences available for comparison.

\subsection{How future HAT research can use the mechanism chain}

Before data collection, researchers can use the chain to specify the social unit and outcome level held constant across AI conditions. An experiment on trust should name whether the referent is the AI, another person, the mixed team, or the deploying institution, and report who sees and can contest the AI output. A study of collective intelligence should observe how contributions are combined across members. Workflow research should separate task speed from coordination cost and handoff quality. Accountability research should place visibility beside authority, control, and consequences.

Reporting visibility is the most immediately actionable of these. Naming the audience of an AI output costs a clause, and only 20 of 86 records in the audit supplied an audience-specific description. A study that states whether the output was seen by one participant, by the group, or by someone with evaluative authority makes its coordination claim checkable, and lets a later review place the finding without inferring the relation from the AI's name.

The same structure can guide synthesis and dataset design. Reviews can group findings by mechanism path before comparing effect directions. Benchmarks can annotate AI role together with output visibility, override rights, responsibility, and outcome level. New HAT taxonomies can then describe both the configuration and the process it permits. These uses turn the human arrangement from background context into part of the construct being measured.

\subsection{Synthesis as accountable evidence work}

Recovering these relations makes literature synthesis a form of accountable evidence work and, in its own right, a cooperative-work problem. Review teams coordinate retrieval, interpretation, normalization, and adjudication across records that expose different parts of a social arrangement. The mechanism chain serves as a shared object for that work: a coder must connect a role to a relation and process before assigning the scope of an outcome claim. Computational checks can identify surface-near records and schema problems, but the four-run consensus result shows why agreement alone cannot authorize a denominator. The judgment remains accountable to a text location and an explicit decision rule.

\section{Limitations and Future Work}

Because the purposive sample places different human arrangements under the same audit, its counts describe this 86-paper set. They identify comparison failures and mechanism differences; they do not estimate how frequently each arrangement occurs across HAT research. A prevalence study would need a representative or exhaustively screened corpus with controlled exclusion reasons for the records outside the audit.

Coder overlap covers 24 papers, while the remaining records contribute primary coding and evidence-location notes. The agreement statistics therefore characterize the overlap subset rather than the entire corpus. Final counts come from the adjudicated dataset and should be interpreted as a transparent application of this codebook, not as an estimate of universal coding reliability.

The surface-triage diagnostics carry several limits. Both language models may have encountered these papers during training, an advantage the TF--IDF representation cannot have; the comparison therefore describes performance on this audit rather than isolating what abstract text alone carries. Shared training exposure, model behavior, and omissions in published metadata could all contribute to the cross-family convergence. Abstract availability is also confounded with record provenance, since 26 of 29 metadata-expansion records carried abstracts against 14 of 57 seed records, and expansion records were selected to fill underrepresented arrangement buckets. The with-abstract and without-abstract contrast cannot be read as an effect of the abstract. Each condition was run once per model using interface defaults, and two model families do not establish that the pattern is universal. The model results are supporting diagnostics, not the paper's primary empirical contribution.

Outcome level was assigned by keyword rule over coder-extracted outcome and claim terms. The rule is transparent and re-runnable, but it inherits imprecision in that field and cannot distinguish a construct that a paper measured from one it mentioned or interpreted. Records with several terms receive several levels, so the counts describe reporting vocabulary rather than measurement prevalence or analytic focus.

The five mechanism paths organize recurring relations in the audited papers; they are not an exhaustive taxonomy of HAT or estimates of causal mediation. They were synthesized as nonexclusive analytic routes, so a system can participate in several paths. Future work can apply the chain prospectively and test whether mechanism-matched synthesis explains variation in reported effects better than AI-role or domain labels alone.

\section{Conclusion}

A HAT label and an AI role do not identify a team mechanism. In this audit, full-text reading moved 36 of 74 apparent core candidates out of the core empirical denominator and separated papers that looked similar at screening into evidence about advice, peer coordination, task orchestration, collective facilitation, and oversight. The differences determine whether a claim concerns a person, a team, a workflow, or an institution.

The mismatch begins earlier than synthesis. Fourteen of 27 dyadic records used team or collaboration outcome vocabulary, compared with 20 of 23 multi-human peer-team records, and only 20 of 86 papers specified who could see the AI output (Figure~\ref{fig:evidence-map}). Surface triage reproduced the same collapse. Across two model families, all four runs under-recovered multi-human arrangements relative to dyads; even unanimous decisions diverged from full-text labels in 32\% of screening cases and 34\% of arrangement cases. Consensus over surface evidence was therefore not a substitute for recovering the reported human relation.

The Human-AI Teaming Mechanism Chain connects human and AI roles within an arrangement to relational coupling, team process, and outcome claim. Its practical test is simple: before trust, collective performance, efficiency, or accountability findings are pooled, the relation and process needed by that outcome must remain recoverable. The claim-pooling checkpoint makes that test usable in study design, reporting, review, and benchmark construction.

\bibliographystyle{ACM-Reference-Format}
\bibliography{references}

\clearpage
\appendix
\section{Record-Level Evidence Availability}
\label{app:evidence-map}

\begin{figure}[H]
\centering
\includegraphics[width=\linewidth]{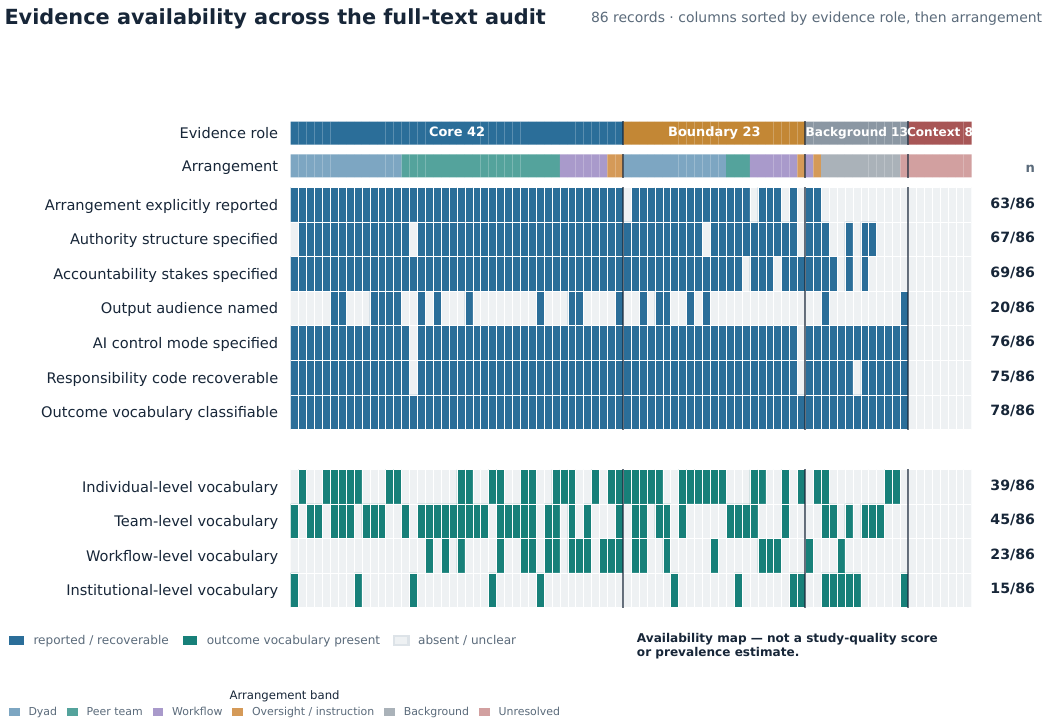}
\caption{Record-level evidence availability across the 86-paper audit. Blue cells mark reported or recoverable analytic fields; teal cells mark the presence of outcome-level vocabulary. Pale cells mark absent, unclear, or unclassifiable fields under the stated rule. Evidence-role and arrangement bands preserve the final audit grouping. Counts at right are audit-set denominators, not prevalence estimates.}
\Description{A matrix with 86 columns, one for each audited paper. The columns are grouped into 42 core empirical, 23 boundary or mechanism, 13 background or theory, and eight context or unresolved records, and are secondarily sorted by arrangement family. Seven blue-and-pale rows show whether arrangement, authority, accountability stakes, output audience, AI control mode, responsibility, and outcome vocabulary were reported or recoverable. Output audience is visibly sparse at 20 of 86 records, while control, responsibility, and outcome vocabulary are recoverable in most core and boundary records. Four teal-and-pale rows show individual, team, workflow, and institutional outcome vocabulary.}
\label{fig:evidence-map}
\end{figure}

Figure~\ref{fig:evidence-map} exposes the reporting substrate behind the aggregate findings without converting missing detail into a study-quality score. Each column is one audited record, sorted first by final evidence role and then by human-arrangement family. The two top bands retain those groupings. The first matrix block shows whether the arrangement was explicitly reported and whether authority, stakes, output audience, AI control, responsibility, and outcome vocabulary were recoverable under the audit rules. The lower block shows the outcome levels identified by the transparent keyword rule described in the analysis; because the coding is multi-label, a record can appear in more than one row.

The map makes two analytic boundaries visible. First, core and boundary evidence usually provides enough information to recover arrangement, control, responsibility, and outcome vocabulary, but audience-specific visibility remains sparse across both roles. Second, background and context records can still supply theory or vocabulary while lacking the relational detail needed for empirical claim pooling. The complete audit table retains the record identifiers, source-location notes, and original field values needed to inspect any column.

\end{document}